\documentclass[pdflatex,sn-basic]{sn-jnl}

\usepackage[utf8]{inputenc}
\usepackage{graphicx}
\usepackage{amsmath,amssymb}
\usepackage{booktabs}
\usepackage{array}
\usepackage{tabularx}
\usepackage{ragged2e}
\usepackage{xcolor}

\newcolumntype{L}[1]{>{\RaggedRight\arraybackslash\hspace{0pt}}p{#1}}
\newcolumntype{Y}{>{\RaggedRight\arraybackslash\hspace{0pt}}X}

\begin{document}

\title[Hybrid Pauli-Correlator Architecture]{Hybrid Quantum--Classical Pauli-Correlator Architecture for Image Reconstruction}

\author[1]{\fnm{Sparsho} \sur{Chakraborty}}\email{25ph05023@iitbbs.ac.in}

\author*[2]{\fnm{Siddhartha} \sur{Patra}}\email{siddhartha.patra@tcgcrest.org}

\affil[1]{\orgdiv{School of Basic Sciences, Department of Physics}, \orgname{Indian Institute of Technology Bhubaneswar}, \orgaddress{\state{Odisha}, \country{India}}}

\affil*[2]{\orgdiv{Centre for Quantum Engineering, Research and Education}, \orgname{TCG CREST}, \orgaddress{\city{Kolkata}, \country{India}}}

\abstract{We introduce the Pauli-Correlator Autoencoder (PCAE), a hybrid quantum--classical architecture for image reconstruction.
PCAE splits an image into patches and compresses each patch as a matrix product state (MPS).
A small quantum circuit turns each compressed patch into a few measured numbers, and a classical neural network rebuilds the patch from these numbers and the patch position.
A learnable energy term, taken from physics, is added to training and tracked as training proceeds.
Two versions place the six qubits on a line (PCAE1D) or on a grid (PCAE2D).
An averaging step makes the PCAE2D features follow rotations and reflections of the image exactly.
Trained on one image at a time, PCAE2D rebuilds images from six public datasets with a peak signal-to-noise ratio of 25.8 to 41.6 decibels.
On CIFAR-10 images not used in training, a formula-based stand-in for the PCAE2D features performs within 1 decibel of several simpler classical methods.
Five trained models also run on real IBM quantum computers without retraining, and rebuild images at 25.90 to 31.52 decibels.
A short argument and a synthetic test show that, with a linear final layer, measuring qubits in pairs is needed for targets built from products of inputs.
These results show that PCAE works as a hybrid model that rebuilds varied images and runs on today's quantum hardware at the six-qubit scale.}


\maketitle

\section{Introduction}
\label{sec:intro}

Hybrid quantum--classical models are now often proposed for vision tasks.
The aim is to obtain representations with a favourable inductive bias.
Such models can use entanglement, physically motivated regularisers, or tensor-network structure, which standard convolutional networks do not have \citep{Cerezo2021, Tilly2022, Preskill2018}.
The variational quantum eigensolver and its later variants use the fact that physical Hamiltonians split into local terms.
This allows them to build shallow circuits suited to the problem \citep{Cerezo2021, Tilly2022, Bauer2020}.
Quantum image processing gives several ways to store pixel values directly in qubits.
These range from amplitude-style encodings \citep{Le2011} to basis-encoded pixel values \citep{Zhang2013}.
Basic image operations on these encodings have also been developed \citep{Chakraborty2018, Zhou2017, Yao2017}.
Tensor-network methods work well as compressed representations of structured classical data, such as images \citep{Cirac2021, Latorre2005, Han2018, Stoudenmire2016, Ran2020TNCS, Guala2023} and particle-physics events \citep{Araz2022}.
Hybrid models have also been applied to related imaging tasks, such as tomographic reconstruction \citep{Haga2024, Nau2023} and ghost imaging \citep{Zhai2025}.
These solve inverse problems from indirect measurements.
PCAE instead encodes and decodes a known image.
Quantum convolutional and capsule networks are further examples \citep{Slabbert2024, Cong2019, Hur2022, Liu2022}.
PCAE focuses on the combination of a Pauli-correlator latent space, enforced group-equivariant pooling, and a physically motivated energy term.

We introduce the Pauli-Correlator Autoencoder (PCAE1D/PCAE2D), which combines these ingredients into a single architecture with three properties studied jointly.
The first is a \emph{Pauli-correlator latent space}, in which measured VQC observables rather than a raw quantum state form the representation handed to a classical decoder, following quantum-enhanced feature maps \citep{Havlicek2019}.
The second is an \emph{exact, numerically verified $D_4$-equivariant observable-pooling construction}, which group-averages the observable grid over the eight symmetries of the $4\times4$ patch grid rather than motivating symmetry through architecture choices alone.
This follows group-equivariant classical vision \citep{Cohen2016, Bronstein2021} and the equivariant-quantum-neural-network line that formalises symmetry-embedding constructions for variational circuits \citep{Meyer2023, Nguyen2024EQNN, Schatzki2024, West2024, West2024Erratum}.
The third is a \emph{learnable Hamiltonian energy term}, used as an auxiliary training objective and read out as a descriptive diagnostic, with full nearest-neighbour $XX/YY/ZZ$ sectors in PCAE1D and grid-edge $ZZ$ terms in PCAE2D.
Table~\ref{tab:differentiation} contrasts these properties against classical, adversarial, and standard quantum autoencoder baselines.

\subsection*{Novelty relative to the nearest hybrid-vision architectures}

PCAE's closest architectural relatives are three hybrid quantum-vision patterns already established individually in the literature.

\emph{Register-based quantum autoencoders} \citep{Romero2017, WangQAECompression2024, QAEClassification2025} encode an image's pixel amplitudes directly into a qubit register and train a trash-qubit fidelity objective.
PCAE feeds the circuit only quantum-inspired, pre-compressed MPS patch features, not raw pixel amplitudes.
It hands the decoder measured Pauli correlators rather than register amplitudes or a fidelity score.
PCAE is an autoencoder in the usual encoder--decoder sense.
It maps each image patch to a measured observable vector and decodes it back to pixels.
It does not compress a quantum state.
Closer to PCAE, a hybrid autoencoder has placed a circuit between a classical encoder and decoder and passed its per-qubit $\langle Z\rangle$ values to the decoder as the latent vector \citep{Sakhnenko2022}.
However, it uses neither MPS input nor pairwise correlators.

\emph{Quantum-kernel vision models} \citep{Havlicek2019, QuantumKernelDefect2022} embed an input into a quantum feature space and read out a scalar kernel value consumed by a classical kernel method.
PCAE reads out an explicit vector of local and pairwise Pauli expectation values as a latent representation for a trained decoder, and adds enforced group-equivariant pooling and a Hamiltonian energy term that a kernel construction has no analogue for.

\emph{Tensor-network (TN) circuit classifiers} \citep{Guala2023, Stoudenmire2016, Han2018} use MPS structure either as a standalone classical model or as a template for a shallow quantum circuit trained for classification.
PCAE uses MPS compression only as a quantum-inspired \emph{preprocessing} step that sets a separately trained VQC's rotation angles, and targets per-image reconstruction rather than classification.
Further, an MPS feature extractor has been placed before a VQC and trained end to end with it for classification \citep{Chen2021MPSVQC}.
The MPS in PCAE is instead a fixed SVD compression.

PCAE's contribution is the assembly of MPS preprocessing and the three properties above in one pipeline (Table~\ref{tab:differentiation}).
Section~\ref{sec:representativeness} examines how the individual ingredients recur across this literature.

\subsection*{Contributions}

PCAE1D/PCAE2D combines MPS patch features, a six-qubit VQC, a 19--27-dimensional Pauli-correlator vector, 2D Fourier positional encoding, a graph-Hamiltonian energy term, and a classical decoder.
An exact, numerically verified $D_4$-equivariant observable-pooling construction makes the PCAE2D observable map equivariant to the eight symmetries of the square, to numerical precision (Section~\ref{sec:d4}).
The guarantee follows from Reynolds group averaging over the $D_4$ orbit \citep{Puny2022} rather than from anything quantum, and the same construction gives the same exactness for a purely classical feature map.
We report it as a design property of the pooled observable map.

Identically configured per-image fitting experiments cover six real image datasets (Section~\ref{sec:sameset_results}).
On held-out CIFAR-10, a closed-form surrogate of the PCAE2D observable map meets the $\pm1$~dB equivalence criterion against five of seven resource-matched classical/ablated feature maps, under two one-sided tests (TOST) with Holm correction.
It clearly outperforms the two random-feature null controls (Section~\ref{sec:ablation_relationship}).
Five fitted checkpoints are replayed on IBM hardware without retraining and decoded from hardware-measured observables (Section~\ref{sec:hardware_pilot}).
A noise/shot-budget sweep and repeated execution across backends support the pilot (Sections~\ref{sec:hardware_robustness}--\ref{sec:hardware_anchors}).
In a leakage-audited protocol with a fixed linear readout, pair observables are necessary when the target depends explicitly on products of input pairs (Section~\ref{sec:entanglement_necessity}).

The appendices give the training settings, the matched-budget ablation, further held-out results and the hardware methodology.

\section{Pauli-Correlator Autoencoder: Architecture}
\label{sec:architecture}

PCAE treats each spatial patch as a finite spin state and compresses it as a matrix product state (MPS).
A variational quantum circuit (VQC) then measures a fixed set of Pauli observables, which a classical multi-layer perceptron (MLP) decodes back into pixels.
A learnable graph-Hamiltonian energy term regularises training.
Parameterised Hamiltonian learning approaches treat the Hamiltonian as the primary learned object \citep{Shi2022}.
In PCAE it is instead an auxiliary objective alongside the reconstruction loss, and its learned energy is read out as a diagnostic.

\paragraph{Problem statement.}
Given an image $x$ partitioned into non-overlapping patches $\{x_p\}_{p=1}^{P}$, PCAE learns a per-patch reconstruction map.
Each patch is compressed to a quantum-inspired MPS feature vector $\phi_{\mathrm{MPS}}(x_p)$, a learned linear layer sets the VQC rotation angles, and measuring a fixed set of Pauli observables yields the correlator feature map $\Phi(x_p)\in\mathbb{R}^{d}$ ($d=27$ for PCAE1D, $19$ for PCAE2D).
With $n=6$ qubits, bond set $\mathcal{E}$, and $c$ measured pair channels per bond,
\begin{equation}
  \label{eq:obs_dim}
  d=2n+c\lvert\mathcal{E}\rvert=
  \begin{cases}
    2(6)+3(5)=27, & \text{PCAE1D ($ZZ,XX,YY$ on a 5-bond chain)},\\
    2(6)+1(7)=19, & \text{PCAE2D ($ZZ$ on 7 grid edges)},\\
    2(6)+1(5)=17, & \text{$ZZ$-only ablation control}.
  \end{cases}
\end{equation}
All PCAE1D and PCAE2D circuit results use $d=27$ and $d=19$.
A classical decoder $D_\theta$ maps $\Phi(x_p)$, together with a Fourier positional encoding $\pi(p)$, back to a reconstructed patch $\hat{x}_p=D_\theta\!\big(\Phi(x_p),\pi(p)\big)$.
Training minimises the pixel-reconstruction loss with the graph-Hamiltonian energy term $E(\cdot)$ as an auxiliary regulariser,
\begin{equation}
  \label{eq:objective}
  \mathcal{L}(\theta)=\frac{1}{Pm}\sum_{p=1}^{P}\big\|x_p-\hat{x}_p\big\|_2^{2}
  \;+\;\lambda\,\frac{1}{P}\sum_{p=1}^{P}E\big(\Phi(x_p)\big),
\end{equation}
where $m$ is the number of pixels per patch, so the first term is the mean squared pixel error, and the second term is the mean graph-Hamiltonian energy over patches and $\lambda$ sets its weight (Section~\ref{sec:regularizer_inert} reports its effect).
The energy of patch $p$ is
\begin{equation}
  \label{eq:energy}
  E\big(\Phi(x_p)\big)=\sum_{(u,v)\in\mathcal{E}}\;\sum_{\alpha\in S}
  J^{\alpha}_{uv}\,\langle\sigma^{\alpha}_{u}\sigma^{\alpha}_{v}\rangle_{p},
\end{equation}
where $\mathcal{E}$ is the bond set of Eq.~(\ref{eq:obs_dim}), $S=\{x,y,z\}$ for PCAE1D and $S=\{z\}$ for PCAE2D, and the couplings $J^{\alpha}_{uv}$ are learned.
The shared pipeline is
\begin{equation}
\begin{aligned}
\text{Image}
  &\rightarrow \{\text{Patches}\} \rightarrow \text{MPS}\\
  &\rightarrow \text{46-D/22-D Feature} \rightarrow \text{6-D Vector}\\
  &\rightarrow \text{27-D/19-D Latent} \rightarrow \text{Decoder}\\
  &\rightarrow \text{Reconstruction}.
\end{aligned}
\end{equation}

\paragraph{Why MPS features.}
Treated as an amplitude-encoded quantum state, a patch of $n$ sites has a coefficient tensor with $2^{n}$ entries.
Matrix-product-state factorisation instead represents it as a chain of small tensors bounded by a fixed bond dimension $\chi$.
This is a compact, singular-value-controlled compression whose entanglement-entropy profile is already known to track image compressibility \citep{Latorre2005, Stoudenmire2016}, and whose local expectations and correlations are cheap features that do not require a full state-preparation circuit.

\paragraph{Why a Pauli-correlator latent space rather than a raw quantum state.}
Handing the classical decoder a fixed vector of measured Pauli expectation values, instead of full state tomography, keeps the quantum-to-classical interface at a size linear in the qubit count, on the bounded-degree graphs used here, rather than exponential.
The execution cost also grows with the number of measurement settings and the shots per setting.
It also follows the measured-observable-as-feature pattern used by quantum-enhanced feature maps generally \citep{Havlicek2019} and by projected quantum kernels built from local reduced density matrices \citep{Huang2021PowerData}.
This choice is what makes the energy diagnostics of Section~\ref{sec:regularizer_inert} and the training-dynamics diagnostics of Section~\ref{sec:phase_transition} possible, since a full state vector would not expose a comparably compact, physically meaningful summary.

\paragraph{Why a grid rather than a chain interaction topology for PCAE2D.}
Arranging the six qubits on a $2\times3$ grid, rather than a linear chain, gives the latent a two-dimensional coupling pattern.
Each patch is processed by its own circuit, so the grid couples the six circuit inputs of one patch, not neighbouring patches.
The $D_4$ pooling of Section~\ref{sec:d4} does not depend on this choice.
It acts by permuting the $4\times4$ grid of patches in an image, and a $2\times3$ graph has no $90^\circ$ rotation symmetry of its own.
PCAE1D's chain topology is the control without a grid.
It is used to measure what the grid adds (Section~\ref{sec:topology}).

\begin{table}[t]
	\centering
	\caption{The two executed configurations.
	Both use six qubits and the same decoder.
	They differ in patch size and in the measured observable set.}
	\label{tab:configurations}
	\begin{tabular}{@{}l c c@{}}
		\toprule
		& PCAE1D                 & PCAE2D                  \\
		\midrule
		Benchmark                     & \textit{Monalisa}     & multi-dataset          \\
		Image size                    & $256\times256$        & $32\times32$ (resized)  \\
		Patch size                    & $64\times64$          & $8\times8$             \\
		MPS sites                     & 12                    & 6                      \\
		Local $X/Z$ expectations      & 24                    & 12                     \\
		Nearest-neighbour $ZZ$        & 11                    & 5                      \\
		 Bipartite entropy features           & 11                   & 5                      \\
		MPS feature vector            & 46-D                  & 22-D                   \\
		\midrule
		Measured observables          & 27-D                  & 19-D                   \\
		\quad local $Z$, local $X$    & 6, 6                  & 6, 6                   \\
		\quad pair terms              & 5 each $ZZ,XX,YY$     & 7 grid-edge $ZZ$       \\
		Fourier positional vector     & 8-D                   & 4-D                    \\
		\bottomrule
	\end{tabular}
\end{table}

\begin{figure}[t]
  \centering
  \includegraphics[width=\textwidth]{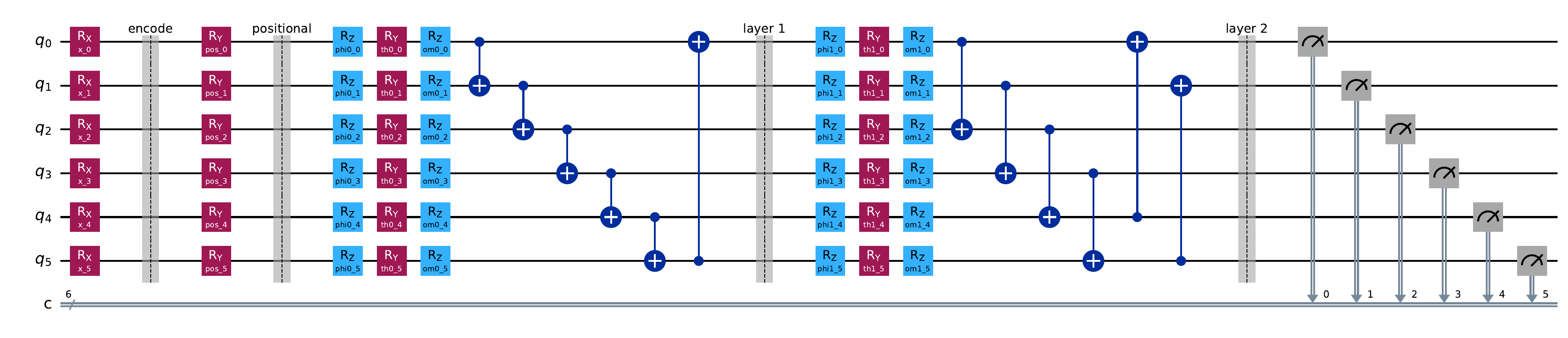}\\[6pt]
  \includegraphics[width=\textwidth]{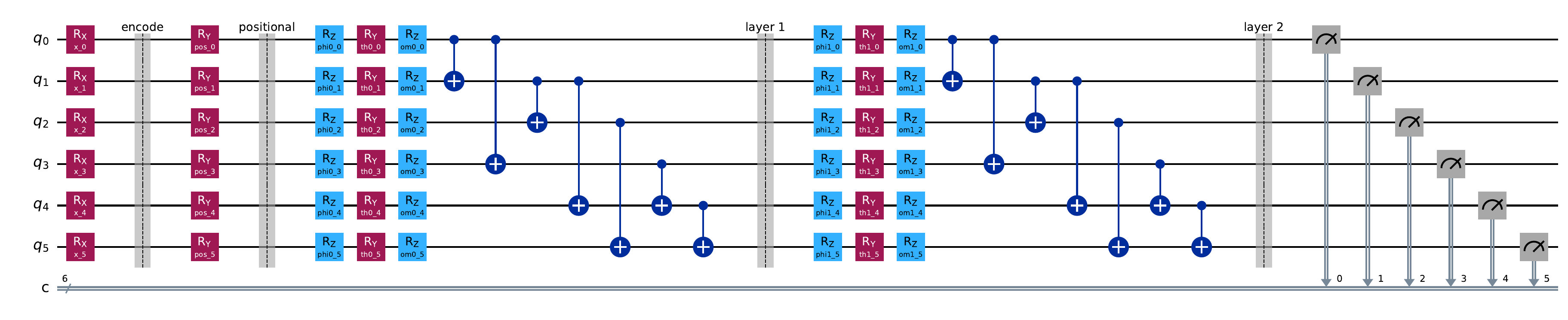}
  \caption{The PCAE1D (top) and PCAE2D (bottom) variational ans\"atze, as built for execution, before transpilation (encoding $R_X$ rotations, positional $R_Y$ rotations, then two entangling layers of $R_Z$-$R_Y$-$R_Z$ single-qubit rotations with a CNOT ring for PCAE1D or fixed grid-edge CNOTs for PCAE2D).
  Both circuits were rebuilt gate-for-gate in Qiskit and validated against the original PennyLane training circuit before use in the hardware pilot of Section~\ref{sec:hardware_pilot}.}
  \label{fig:hvk_ansatz}
\end{figure}

Two preprocessing configurations were executed, one per topology.
Both follow the same three stages.
Each image is first divided into sixteen non-overlapping patches, except in the topology comparison (Appendix~\ref{sec:training_settings}).
Every patch is then $\ell_2$-normalised and sequentially SVD-compressed to a matrix product state of bond dimension $\chi=4$.
Finally, the local expectations, nearest-neighbour correlations and bipartite entropy features of that state are read off as the quantum-inspired MPS feature vector.
The two configurations differ only in the sizes involved, which Table~\ref{tab:configurations} lists in full.
The $2^{n}$ pixel values of a patch ($n=6$ for $8\times8$, $n=12$ for $64\times64$) are flattened in row-major order and reshaped into an order-$n$ tensor with binary indices, so each index is one site of dimension~2.
The tensor is compressed with the \texttt{compress} method of quimb \citep{Gray2018} at maximum bond dimension $\chi$.
With its default settings, the compressed state is in right-canonical form, with the orthogonality centre on the first site.
The features are expectation values and Schmidt spectra of the compressed state, so they do not depend on its canonical form.
The entropy feature of each cut is computed from the Schmidt probabilities $p_k$ of that cut, which are the squared singular values.
The implementation squares them a second time and renormalises them, $q_k=p_k^2/\sum_j p_j^2$, and returns $-\sum_k q_k\ln q_k$.
This feature agrees with the von Neumann entropy $-\sum_k p_k\ln p_k$ for a product cut and for a flat Schmidt spectrum, but not in general.
All models in this paper were trained with this feature.
In the multi-dataset runs of Section~\ref{sec:sameset_results}, $10^{-4}$ is added to every pixel before normalisation, so that all-zero patches, such as MNIST background, can be normalised.
The offset enters only the MPS input.
The decoder is still trained to reconstruct the unmodified patch.
The remaining stages are common to both.
A learned linear layer compresses the MPS feature vector to six circuit angles, one per qubit.
After the circuit is executed, a two-hidden-layer MLP (128 and 256 units) maps the measured observables together with the Fourier positional vector back to a single greyscale patch.
The positional encoding follows established use as a spatial inductive bias in classical generative vision models \citep{Xu2021}.
Fourier features of this kind also help coordinate-based networks fit fine image detail \citep{Tancik2020}.
During training the circuit is simulated exactly.
Unless otherwise stated, training used additive zero-mean Gaussian noise with standard deviation $0.01$ on the observable vector.
Artificial Gaussian noise was not added during evaluation or hardware replay.
Finite-shot estimates enter only in the hardware replay and the noisy simulation of Sections~\ref{sec:hardware_pilot}--\ref{sec:hardware_anchors}.
Figure~\ref{fig:hvk_ansatz} shows the exact PCAE1D and PCAE2D ansatz circuits.
These were rebuilt in Qiskit for the hardware pilot of Section~\ref{sec:hardware_pilot}, and numerically verified against the training-time PennyLane circuit before use.

Table~\ref{tab:variants} summarises the tested variants.
PCAE1D uses a linear 6-qubit chain (5 nearest-neighbour bonds, 27-D observable vector).
PCAE2D arranges the same 6 qubits on a $2\times3$ lattice (19-D latent signature).
A $D_4$-pooled PCAE2D variant (Section~\ref{sec:symmetry}) group-averages the observable grid over the eight symmetries of the $4\times4$ patch grid, $\Phi_{D_4}(x)=\tfrac{1}{8}\sum_{g\in D_4}P_g^{-1}\Phi(gx)$, which is equivariant by construction.
Each circuit has two entangling layers, with 12 CNOTs for PCAE1D and 14 for PCAE2D before transpilation (Figure~\ref{fig:hvk_ansatz}).

\begin{table}[t]
  \centering
  \caption{PCAE variants tested in this paper.}
  \label{tab:variants}
  \begin{tabular}{@{}l l c l@{}}
    \toprule
    Variant                 & Topology                  & Latent dim. & Symmetry \\
    \midrule
    PCAE1D                   & 6-qubit chain             & 27-D        & Unpooled            \\
    PCAE2D                   & $2\times3$ grid           & 19-D        & Unpooled      \\
    $D_4$-pooled PCAE2D      & $2\times3$ grid, pooled   & 19-D        & $D_4$-equivariant (exact) \\
    \bottomrule
  \end{tabular}
\end{table}

\section{Experimental Setup}
\label{sec:protocol}

The reconstruction results in Sections~\ref{sec:sameset_results}, \ref{sec:no_generalization} and \ref{sec:hardware_pilot}--\ref{sec:hardware_anchors} use single-image fitting, the protocol developed for PCAE on \textit{Monalisa}.
A fresh model is optimised directly on its target image, with no held-out split.
These results therefore measure per-image fitting and reconstruction capability.
The hardware pilot replays the resulting fitted checkpoints on real devices without retraining.
The symmetry and restricted pair-observable diagnostics use closed-form feature maps and the separate protocols stated in their respective sections.
The training-dynamics readouts of Section~\ref{sec:phase_transition} are reported as a descriptive diagnostic, with their scope set by the negative control given there.
Held-out generalisation and component necessity are evaluated under the resource-matched statistical protocol, using a closed-form surrogate of the PCAE2D observable map (Section~\ref{sec:ablation_relationship}).
All settings report MSE and PSNR $=20\log_{10}(1/\sqrt{\mathrm{MSE}})$.
Image settings additionally report SSIM.
Gradient-trained PCAE models were optimised using full-batch Adam, with learning rates $3\times10^{-3}$ for PCAE1D and $4\times10^{-3}$ for PCAE2D, with Hamiltonian-loss weight $\lambda=0.01$.
We used 200 optimisation steps for the \textit{Monalisa} and CIFAR-10 hardware-checkpoint fits, 200 steps for the six-dataset same-set fits, and 90 steps for the topology comparison.
The \textit{Monalisa} run used seed 42 and the topology comparison used seeds $\{0,1,2\}$.
Circuit rotation parameters were initialised independently from $\mathcal U(0,\pi)$, and Hamiltonian couplings from $0.1\,\mathcal N(0,1)$.
The linear projection and decoder layers used PyTorch's default initialisation.
The software environment and the per-experiment settings are given in Appendix~\ref{sec:training_settings}.

\section{Results}
\label{sec:results}

\subsection{Reconstruction across six real image datasets}
\label{sec:sameset_results}

For each image, we trained a fresh PCAE2D model using non-overlapping $8\times8$ patches (16 patches/image, 200 optimisation steps).
The sample covers CIFAR-10 \citep{Krizhevsky2009}, MNIST \citep{LeCun1998}, Fashion-MNIST \citep{Xiao2017}, and the PathMNIST, BloodMNIST, and PneumoniaMNIST subsets of MedMNIST~v2 \citep{Yang2023}.
Table~\ref{tab:sameset_multi_dataset} reports mean PSNR between $25.8$ and $41.6$~dB (SSIM $0.91$--$1.00$).
The architecture trains end to end and fits varied image structures under a common per-image protocol.
A position-only ablation of the same decoder also fits these sampled images (Table~\ref{tab:sameset_multi_dataset}).
It has a higher mean PSNR on three of the six datasets, and is more than 5~dB higher on CIFAR-10 and PneumoniaMNIST.
Table~\ref{tab:sameset_multi_dataset} establishes fitting feasibility.
Evaluation-time substitutions show that trained models are sensitive to their observable inputs (Section~\ref{sec:hardware_pilot}).
The $D_4$ construction and the restricted pair-observable diagnostic establish separate properties of the representation (Sections~\ref{sec:d4} and~\ref{sec:entanglement_necessity}).
This protocol is per-image fitting, in the same sense as implicit neural representations, which fit one network to one image or scene \citep{Ulyanov2020DIP, Sitzmann2020, Tancik2020, Mildenhall2020NeRF, Dupont2021COIN, Strumpler2022}.
Quantum versions of this idea fit a single signal or image with a variational circuit \citep{Zhao2024QIREN, Fujihashi2024QINC, LizzioBosco2026QISR}.
The closest of these, QIREN, is a quantum coordinate network aimed at a fitting advantage over classical Fourier networks \citep{Zhao2024QIREN}.
PCAE is instead a patch autoencoder with a measured-correlator bottleneck, and it aims at a practical hybrid pipeline whose quantum part is the measured observable vector.
Generalisation to held-out images is tested in Section~\ref{sec:ablation_relationship}.
A compression comparison would also need a bit rate \citep{Dupont2021COIN, Strumpler2022}, so we report fitting quality here.

\begin{table}[t]
  \centering
  \caption{PCAE2D same-set reconstruction across six real image datasets (per-image trained, 100 steps, mean $\pm$ std over sampled images).
  SSIM is computed globally over each image.
  The position-only ablation is the same decoder trained with all 19 observable inputs set to zero, on the same images and patch layout, with the same learning rate and 100-step budget.
  It uses seed 0 for every image.
  The PCAE2D seeds were not recorded, so the two are not paired by initialisation.}
  \label{tab:sameset_multi_dataset}
  \small
  \setlength{\tabcolsep}{4pt}
  \begin{tabular}{@{}l c r@{\,$\pm$\,}l r@{\,$\pm$\,}l r@{\,$\pm$\,}l r@{\,$\pm$\,}l@{}}
    \toprule
                     &            & \multicolumn{4}{c}{PCAE2D} & \multicolumn{4}{c}{Position-only ablation} \\
    \cmidrule(lr){3-6}\cmidrule(l){7-10}
    Dataset          & $n$ images & \multicolumn{2}{c}{PSNR (dB)} & \multicolumn{2}{c}{SSIM} & \multicolumn{2}{c}{PSNR (dB)} & \multicolumn{2}{c}{SSIM} \\
    \midrule
    CIFAR-10         & 3 & 37.75 & 2.38 & 0.9986 & 0.0003 & 42.98 & 8.54 & 0.9991 & 0.0008 \\
    MNIST            & 3 & 25.77 & 2.23 & 0.9783 & 0.0116 & 25.95 & 2.31 & 0.9796 & 0.0101 \\
    Fashion-MNIST    & 3 & 34.55 & 3.05 & 0.9980 & 0.0006 & 33.70 & 0.49 & 0.9968 & 0.0026 \\
    PathMNIST        & 3 & 36.86 & 2.11 & 0.9118 & 0.0627 & 35.49 & 1.92 & 0.8904 & 0.0655 \\
    BloodMNIST       & 2 & 36.57 & 2.36 & 0.9922 & 0.0063 & 34.67 & 3.98 & 0.9833 & 0.0151 \\
    PneumoniaMNIST   & 2 & 41.60 & 0.98 & 0.9985 & 0.0006 & 47.28 & 1.04 & 0.9996 & 0.0002 \\
    \bottomrule
  \end{tabular}
\end{table}

\subsection{Scope of per-image fitting}
\label{sec:no_generalization}

A model optimised on one image is specific to that image.
In this single-seed test, PCAE2D with $8\times8$ patches is trained on \textit{Monalisa} alone for 200 steps and then evaluated, without further training, on Michelangelo's \textit{Hand of God}.
Zero-shot transfer to \textit{Hand of God} reaches $8.31$~dB PSNR, with SSIM $0.1044$.
Extending training to include \textit{Hand of God} for another 200 steps recovers $28.63$~dB, with SSIM $0.9858$.
Section~\ref{sec:ablation_relationship} tests the held-out case directly.

\subsection{Held-out comparison with a closed-form surrogate}
\label{sec:ablation_relationship}
The results above measure per-image fitting.
A separate question is whether PCAE's feature map improves reconstruction on real, \emph{held-out} images under a resource-matched multi-seed protocol.
This comparison uses a closed-form surrogate of the PCAE2D observable map, so that many seeds and controls can be run.
The surrogate combines 18 local patch statistics, six products of local-feature pairs that stand in for the pairwise $ZZ$ correlators, and eight positional features.
It is specified by hand and computed without running the circuit.
Thus, the term has a different meaning here from the classical surrogates of \citet{Schreiber2023}, which are derived from a trained quantum model.
The main results are summarised below, including the equivalence tests of Table~\ref{tab:tost_main}.
Each of five split seeds fits the readout on six CIFAR-10 images and evaluates it on four disjoint held-out images, so each mean covers 20 held-out reconstructions.

On held-out CIFAR-10, the raw/local linear control reaches $18.80\pm1.42$~dB, against the PCAE2D surrogate's $18.30\pm1.62$~dB.
Aggregating held-out images within each split gives a PCAE2D-minus-control difference of $-0.50$~dB, with bootstrap 95\% CI $[-0.76,-0.29]$~dB and Wilcoxon $p=0.0625$ at $n=5$ seeds \citep{Efron1993, Wilcoxon1945}.
The same numerical direction holds on five further real datasets, which use an earlier feature allocation (Appendix~\ref{app:multidataset}).
The surrogate clearly beats the random-feature controls (Table~\ref{tab:tost_main}), and its mean stays within 1~dB of the simpler maps.
The raw/local linear control is $0.50$~dB ahead of it.

All controls are matched at the readout.
Each passes a 32-D feature vector to the same linear readout with 2112 parameters, on the same splits.
This matches the readout resources.
The controls differ in what their features can express.
The no-entanglement and $ZZ$-only controls remove all or half of the six raw pair products, and the quadratic classical control adds second-order classical terms.
The $ZZ$-only control keeps exactly three of them.
The surrogate is a 32-D map, not the 19-D circuit map of Table~\ref{tab:configurations}.
The quadratic classical control keeps the 18 local features, the first six of its nine products of adjacent local-feature pairs, and the eight positional features, which gives the same 32-D width.
Neither the surrogate nor the quadratic control includes the sines of its products.
The $ZZ$-only and raw/local maps are padded with three and six zero columns, respectively, to reach 32-D.

A TOST equivalence test \citep{Schuirmann1987, Lakens2017} at a declared $\pm1$~dB margin asks whether each gap is small enough to call two maps equivalent rather than merely ``not shown to differ''.
The estimand is the seed-level PCAE2D-minus-control PSNR difference, averaged over held-out images within each seed.
Without correction, equivalence requires both one-sided tests to reject at $\alpha=0.05$.
This is the same as the 90\% confidence interval lying inside $[-1,+1]$~dB (Table~\ref{tab:tost_main}).
The margin is a declared choice.
With five seeds, $p=0.0625$ is the smallest attainable nonzero two-sided Wilcoxon signed-rank $p$-value, so that test cannot reject at the 0.05 level.
The $t$-based 90\% interval is wider than the percentile-bootstrap 95\% interval in these data.
Bootstrap intervals can be unstable with only five seed-level observations.
For scale, a 1~dB PSNR difference corresponds to a factor of approximately $10^{0.1}\simeq1.26$ in MSE.
After Holm correction \citep{Holm1979} across the seven comparisons, five controls meet the equivalence criterion against PCAE2D within that margin.
These are raw/local linear, no entanglement, $ZZ$-only observables, quadratic classical features, and shuffled pair observables.
The raw/local linear control has the largest of these gaps.
It meets the criterion with TOST $p=0.009$ and Holm-adjusted $p=0.027$.
The two weakest controls, strict classical random features and random-VQC, are not equivalent because the surrogate substantially outperforms them.
The trained circuit has been tested on held-out images in the topology comparison of Section~\ref{sec:topology}.
Trained on two images for 90 steps, PCAE2D reaches $11.57\pm1.53$~dB on three unseen CIFAR-10 images, about 1~dB below a matched random-latent control under the same protocol ($12.64\pm2.12$~dB, Appendix~\ref{app:random_latent}).
Held-out generalisation of the trained circuit is therefore not established at this budget.
In a larger end-to-end test with 150 training and 50 held-out images, a parameter-matched classical control reconstructs the held-out images better than the trained PCAE2D model (Appendix~\ref{app:multidataset}).

\begin{table}[t]
  \centering
  \caption[TOST equivalence test on held-out CIFAR-10]{TOST equivalence test on held-out CIFAR-10 ($\pm1$~dB margin, $\alpha=0.05$, five seed-level paired differences, closed-form PCAE2D surrogate minus control).
  The verdict uses the Holm-adjusted $p$.
  Without correction, equivalence is the same as the 90\% CI lying inside $[-1,+1]$~dB.
  ``No entanglement'' removes the pair products.
  ``Random VQC'' is a block of Gaussian random features.}
  \label{tab:tost_main}
  \footnotesize
  \setlength{\tabcolsep}{2pt}
  \begin{tabular}{@{}l r c r l l@{}}
    \toprule
    Control                          & $\Delta$PSNR (dB) & 90\% CI (dB)       & TOST $p$  & Holm $p$ & Verdict \\
    \midrule
    Raw/local linear                 & $-0.50$ & $[-0.78,-0.22]$ & $0.009$   & $0.027$ & Equivalent \\
    No entanglement                  & $-0.16$ & $[-0.52,+0.20]$ & $0.004$   & $0.016$ & Equivalent \\
    $ZZ$-only observables            & $-0.26$ & $[-0.46,-0.07]$ & $0.001$   & $0.004$ & Equivalent \\
    Quadratic classical              & $-0.14$ & $[-0.24,-0.04]$ & $<0.001$   & $<0.001$ & Equivalent \\
    Shuffled pair observables        & $-0.13$ & $[-0.38,+0.11]$ & $0.001$  & $0.004$ & Equivalent \\
    Strict classical random features & $+2.45$ & $[+2.06,+2.84]$ & $0.999$   & $1.000$ & PCAE2D better \\
    Random VQC                       & $+7.15$ & $[+5.73,+8.57]$ & $1.000$   & $1.000$ & PCAE2D better \\
    \bottomrule
  \end{tabular}
\end{table}

A matched-budget, multi-seed component ablation gives a consistent picture on per-image reconstruction.
The ablation uses 240 steps and 3 seeds per variant (4 for the baseline).
It evaluates the stitched image after seam blending, which lowers the PSNR of well-fitted models and compresses the differences between them (Section~\ref{sec:regularizer_inert}).
The quantum baseline, the classical-replacement control and most other variants lie within a fraction of a decibel of one another, so this evaluation does not separate them.
Apart from a freeze-classical sanity check, only a circuit-free random-VQC control performs clearly worse.
Freeze-classical holds the decoder fixed, so it cannot learn to reconstruct and is expected to fail.
The ablation uses a descriptive pooled-standard-deviation comparison rule at 3 seeds rather than a calibrated test.
Full results are in Appendix~\ref{sec:core_multiseed_rerun}.
Section~\ref{sec:dequant} discusses why this result and the pair-observable necessity result of Section~\ref{sec:entanglement_necessity} are consistent.

\subsection{A real hardware reconstruction pilot}
\label{sec:hardware_pilot}

The results above are noiseless-simulator results.
We now report a small but complete image-reconstruction pilot executed on real IBM Quantum hardware (\texttt{ibm\_fez}, a 156-qubit Heron-class processor).

The pilot covers five images.
These are the full 16-patch \textit{Monalisa} benchmark and four native-resolution CIFAR-10 images (\textit{cat}, \textit{ship}$\times2$, \textit{airplane}), each optimised directly on its own target image following the same per-image protocol as Sections~\ref{sec:sameset_results} and~\ref{sec:no_generalization}.
For each image we took an already-trained PCAE1D/PCAE2D checkpoint's exact learned parameters and rebuilt its measurement circuits gate-for-gate in Qiskit.
We verified the translation against the original PennyLane simulator to within numerical noise, at most $7.9\times10^{-4}$ maximum absolute error per observable across every patch.
We then executed the same circuits on real hardware with 256 shots per measurement basis.
This is a fixed per-basis shot budget, rather than an adaptive shadow-tomography-style estimation scheme \citep{Huang2020Shadows}.
The hardware-measured observables were decoded by the same trained classical decoder, with no retraining and no simulator involved anywhere in the reconstruction path.

\begin{figure}[t]
  \centering
  \includegraphics[width=\textwidth]{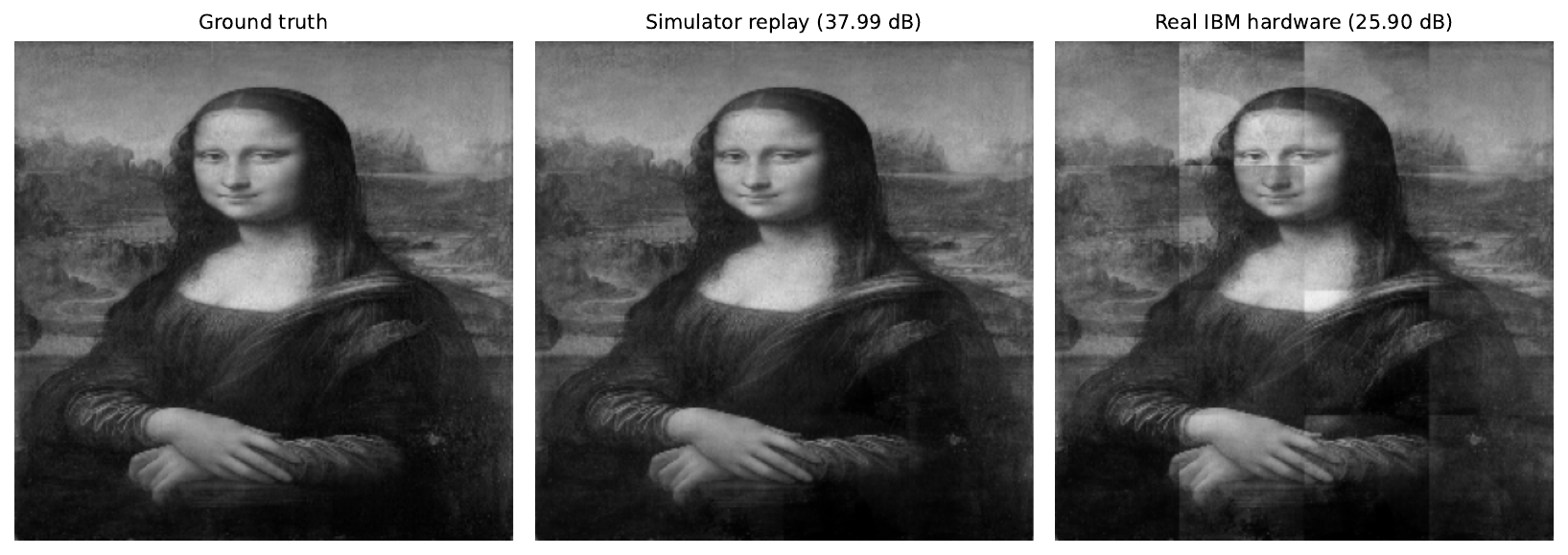}
  \caption{\textit{Monalisa} reconstructed from real IBM Quantum hardware measurements (right) against the same trained checkpoint's noiseless-simulator replay (centre) and ground truth (left).
  The hardware image is visibly noisier and exhibits patch-boundary banding, while the subject remains recognisable.}
  \label{fig:hardware_reconstruction_monalisa}
\end{figure}

\begin{figure}[t]
  \centering
  \includegraphics[width=\textwidth]{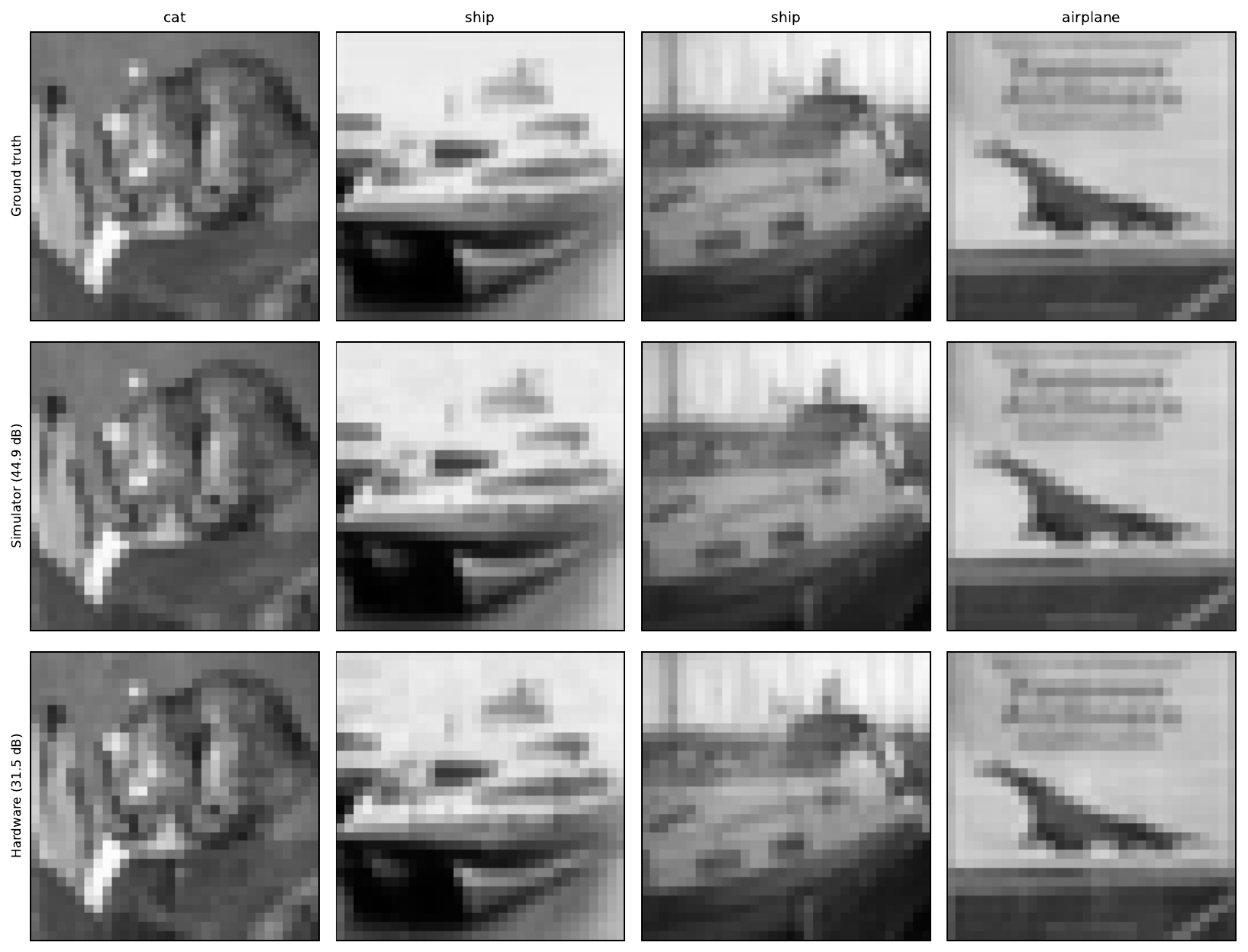}
  \caption{Four CIFAR-10 images reconstructed from real IBM Quantum hardware measurements (bottom row) against noiseless-simulator replays of the same trained PCAE2D checkpoints (middle row) and ground truth (top row).}
  \label{fig:hardware_reconstruction_cifar}
\end{figure}

Reconstructions from real, noisy hardware measurements retain recognisable image structure.
\textit{Monalisa} reaches $25.90$~dB PSNR from hardware measurements, against a $37.99$~dB noiseless-simulator replay of the identical circuit (Figure~\ref{fig:hardware_reconstruction_monalisa}).
The four CIFAR-10 images average $29.56\pm2.03$~dB on hardware against $42.69\pm3.84$~dB in simulation (Figure~\ref{fig:hardware_reconstruction_cifar}, Table~\ref{tab:hardware_pilot_summary}).
The observed gap is consistent with noise in these unmitigated device executions.
Circuits were transpiled at optimisation level 1 and submitted through SamplerV2 with default options.
Dynamical decoupling, Pauli twirling and readout-error mitigation were not enabled \citep[see][]{Temme2017, vandenBerg2022, Kim2023}.
Each image used 256 shots per basis, totalling 12{,}288 shots for \textit{Monalisa} and 8{,}192 for each CIFAR-10 image.
The \textit{Monalisa} hardware reconstruction ($25.90$~dB) remains well above the random-latent control.
On the \textit{Monalisa} hardware checkpoint, substituting random latents for the observables at evaluation gives $12.49$~dB.
This value comes from the training-run evaluation, which blends the seams of the stitched image (Appendix~\ref{sec:hardware_methodology}).
The pilot used 176 circuits.
The contemporaneous account-meter reading increased by 16 quantum-processor seconds.
A similar protocol, with training in simulation and inference on IBM hardware, has been used for MedMNIST classification with error suppression and mitigation \citep{Singh2026}.
Here, it is applied to reconstruction without mitigation.

\begin{table}[t]
  \centering
  \caption{Real IBM hardware reconstruction pilot (backend \texttt{ibm\_fez}, 256 shots/basis, no retraining).
  PSNR compares hardware-decoded and simulator-decoded reconstructions against the same ground-truth image.}
  \label{tab:hardware_pilot_summary}
  \begin{tabular}{@{}l c c c@{}}
    \toprule
    Image                              & Patches & Sim.\ PSNR (dB)   & HW PSNR (dB)      \\
    \midrule
    Monalisa (PCAE1D)                   & 16      & 37.99             & 25.90             \\
    CIFAR cat (PCAE2D)                  & 16      & 44.85             & 31.52             \\
    CIFAR ship, hydrofoil (PCAE2D)      & 16      & 36.56             & 26.44             \\
    CIFAR ship, sea boat (PCAE2D)       & 16      & 42.57             & 31.20             \\
    CIFAR airplane (PCAE2D)             & 16      & 46.79             & 29.10             \\
    \midrule
    CIFAR mean $\pm$ std (4 images)    & ---     & $42.69\pm3.84$    & $29.56\pm2.03$    \\
    \bottomrule
  \end{tabular}
\end{table}

The pilot establishes that the trained PCAE observable channel survives execution on present-day hardware.
The learned parameters, the measurement circuits and the classical decoder all transfer unchanged, and the decoded images remain quantitatively informative at 256 shots per basis on an unmitigated device.
This is the sense in which PCAE runs on real quantum hardware.
The measurement construction and validation results are in Appendix~\ref{sec:hardware_methodology}.

\subsection{Noise/shot trade-off: a calibrated-noise simulation}
\label{sec:hardware_robustness}

The five-image pilot above used a single shot budget (256/basis) and one execution per image, which bounds how much it can say about noise sensitivity on its own.
To use IBM Quantum time well, we separate what a \emph{calibrated device-noise simulation} can establish at zero hardware cost from what genuinely requires hardware.
Using \texttt{qiskit-aer} with the stored calibration snapshot of \texttt{ibm\_fez} (\texttt{FakeFez}, no quantum-processor time consumed), we replay the same already-trained checkpoints' circuits at four shot budgets (256/512/1024/4096 shots per basis, three repeated executions each) and compare against both the noiseless-simulator ceiling and the real-hardware point already reported above.

\begin{figure}[t]
  \centering
  \includegraphics[width=0.7\textwidth]{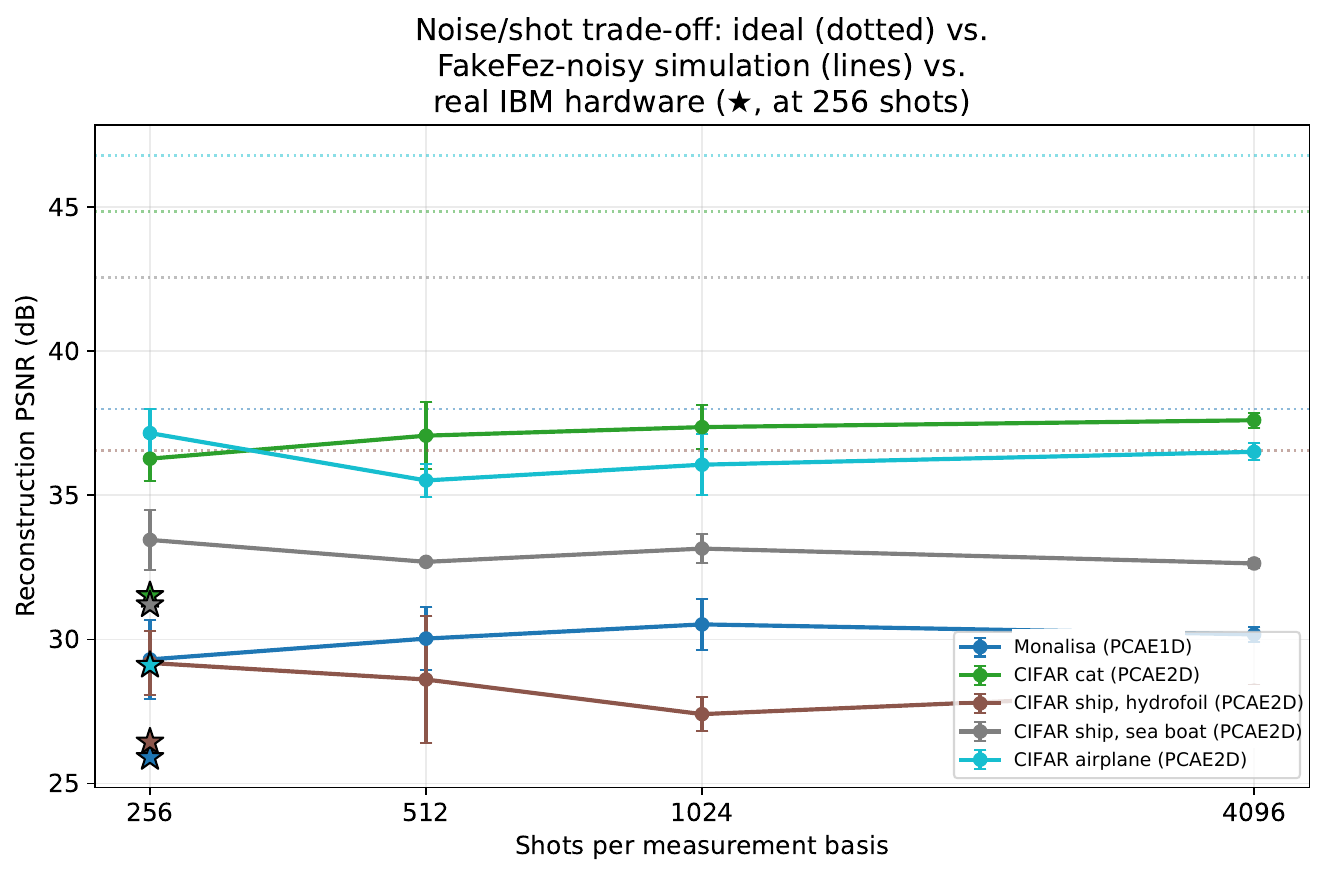}
  \caption{Calibrated device-noise simulation (\texttt{FakeFez}, mean $\pm$ std over three executions).
  Dotted lines show noiseless ceilings and stars show real-hardware measurements.
  Increasing 256 to 4096 shots changes mean PSNR by only $-0.05$~dB.}
  \label{fig:hardware_robustness_shot_sweep}
\end{figure}

Two findings follow, averaged over the five checkpoints (Figure~\ref{fig:hardware_robustness_shot_sweep}).
First, shot noise accounts for only part of the noiseless-to-real gap.
Moving from the noiseless ceiling to a calibrated-noise simulation at the same 256-shot budget already costs $8.68$~dB on average, before any real device is involved.
Second, more shots do not reliably recover it.
Increasing to 4096 shots per basis changes PSNR by only $-0.05$~dB on average across the five checkpoints, with individual images ranging from a $1.33$~dB gain to a $0.96$~dB loss.
At this circuit size and shot range, shot count is not a lever that reliably improves reconstruction quality.

Comparing the 256-shot noisy simulation directly against the real-hardware point isolates what the calibration snapshot does \emph{not} capture.
That comparison leaves a further $4.24$~dB average gap, ranging from $2.26$ to $8.06$~dB across images.
This residual is consistent with effects that the calibration model does not capture, such as drift between calibration and execution, crosstalk, and correlated errors.
The airplane image is the clear outlier, with an $8.06$~dB residual gap, well above any other image.
We report it rather than exclude it, since a single-execution, five-image pilot cannot distinguish an image-specific effect from ordinary job-to-job variation.
This calibrated-noise simulation is a free, repeatable complement to the pilot.
It bounds how much of the observed hardware gap is attributable to modelled device noise rather than unmodelled effects, using zero additional quantum-processor time.

\subsection{Real-hardware anchor points: repeated execution on two further backends}
\label{sec:hardware_anchors}

The simulated trade-off above is useful only if it reflects real devices.
We checked this directly on hardware with four jobs from the free-tier IBM Quantum allowance.
These four jobs used well under $1\%$ of the monthly limit.
Two checkpoints, \textit{Monalisa} and CIFAR \textit{cat}, were replayed with no retraining on \texttt{ibm\_marrakesh} and \texttt{ibm\_kingston}.
Both backends are distinct from the original pilot's \texttt{ibm\_fez}.
We ran them at the original 256-shot budget and, separately, at 1024 shots (Table~\ref{tab:hardware_anchors}).

\begin{table}[t]
  \centering
  \caption{Real-hardware anchor points: same checkpoints, no retraining, on two further backends.
  The original pilot's \texttt{ibm\_fez} point is shown for reference.}
  \label{tab:hardware_anchors}
  \begin{tabular}{@{}l l r r@{}}
    \toprule
    Image       & Backend                       & Shots & PSNR (dB) \\
    \midrule
    Monalisa    & \texttt{ibm\_fez} (original)  &  256  & 25.90     \\
    Monalisa    & \texttt{ibm\_marrakesh}       &  256  & 25.94     \\
    Monalisa    & \texttt{ibm\_kingston}        & 1024  & 26.10     \\
    \addlinespace
    CIFAR cat   & \texttt{ibm\_fez} (original)  &  256  & 31.52     \\
    CIFAR cat   & \texttt{ibm\_kingston}        &  256  & 28.93     \\
    CIFAR cat   & \texttt{ibm\_kingston}        & 1024  & 31.24     \\
    \bottomrule
  \end{tabular}
\end{table}

\textit{Monalisa} reproduces almost exactly across backends at matched shots ($25.94$ vs.\ $25.90$~dB, \texttt{ibm\_marrakesh} vs.\ \texttt{ibm\_fez}), and reaches $26.10$~dB on \texttt{ibm\_kingston} at 1024 shots, consistent with the simulated finding above that shot count is not a reliable lever.
The CIFAR \textit{cat} point is the more informative one.
On \texttt{ibm\_kingston} at the original 256-shot budget it reaches $28.93$~dB, which is $2.59$~dB below the \texttt{ibm\_fez} pilot point.
Its own 1024-shot execution on the same backend recovers to $31.24$~dB, within $0.28$~dB of the original.
This is real job-to-job and backend-to-backend variation of a magnitude comparable to the unmodelled residual identified by the simulator comparison (Section~\ref{sec:hardware_robustness}).
It gives direct evidence that repeated execution matters at least as much as backend choice or shot count.
Repeated runs therefore give more stable estimates of hardware performance.

\subsection{Exact \texorpdfstring{$D_4$}{D4}-equivariant observable pooling}
\label{sec:d4}
\label{sec:symmetry}

\begin{figure}[t]
  \centering
  \includegraphics[width=0.55\textwidth]{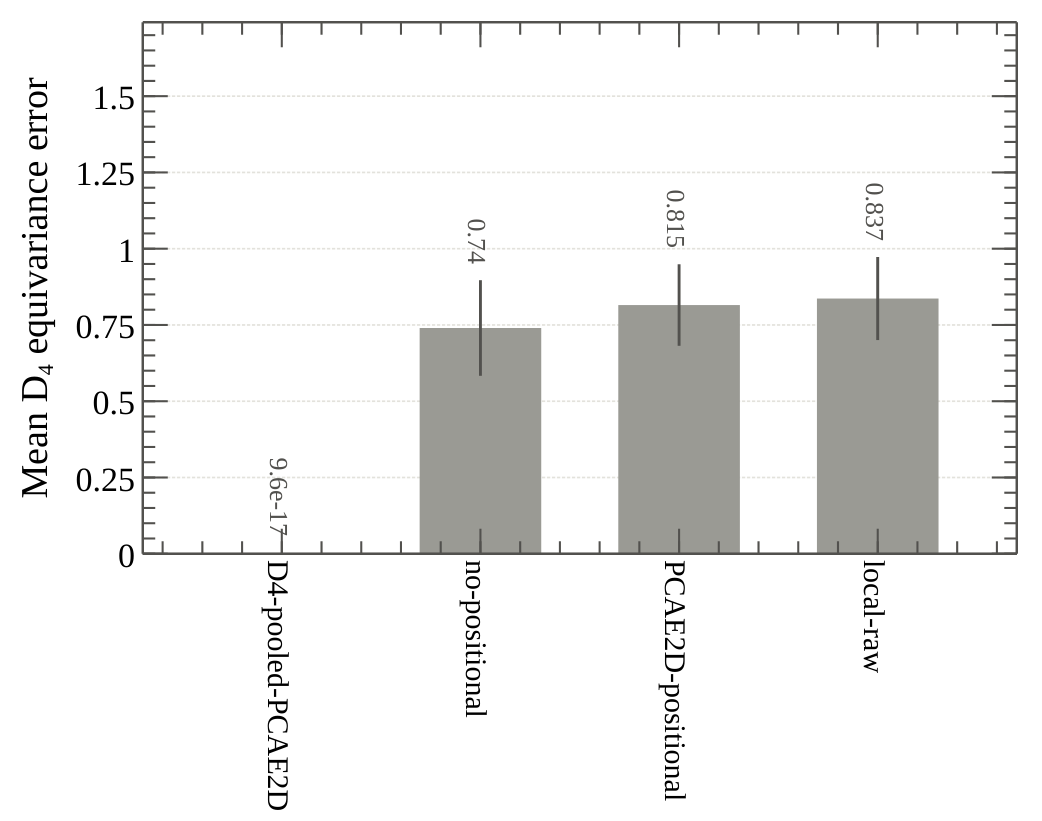}
  \caption[$D_4$ equivariance diagnostic]{$D_4$ equivariance diagnostic.
  The group-pooled map is equivariant to numerical precision.
  The unpooled positional map is not.}
  \label{fig:d4_equivariance}
\end{figure}

A $D_4$-pooled PCAE2D observable-grid map is equivariant to the eight symmetries of the square \emph{by construction}:
\begin{equation}
  \Phi_{D_4}(x)=\frac{1}{8}\sum_{g\in D_4}P_g^{-1}\Phi(gx),
  \qquad
  \Phi_{D_4}(hx)=P_h\,\Phi_{D_4}(x)\ \ \forall\, h\in D_4.
\end{equation}
Here $D_4$ acts on an image $x$ by rotating or reflecting its $4\times4$ grid of patches, and $P_g$ permutes the per-patch observable vectors in the same way, so that $P_{gh}=P_gP_h$.
The property then holds exactly for any base map $\Phi$.
This is the standard group-averaging construction \citep{Puny2022}.
It needs one evaluation of the base map for each of the eight group elements.
Substituting $hx$ and re-indexing the sum with $g'=gh$ gives $P_g^{-1}=P_hP_{g'}^{-1}$, and hence $\Phi_{D_4}(hx)=P_h\Phi_{D_4}(x)$.
The numerical check below therefore tests the implementation, not the property.
We run it on the closed-form surrogate feature map over 7000 patch transforms of cached CIFAR images (Table~\ref{tab:d4_equivariance}, Figure~\ref{fig:d4_equivariance}).
The pooled map's equivariance error is $9.57\times10^{-17}\pm1.26\times10^{-17}$, which is floating-point noise.
The unpooled positional map has an error of $0.815\pm0.129$, and the local/raw control has $0.837\pm0.131$.
Both are of order one.
The patch grid's geometry therefore makes the unpooled map $D_4$-\emph{motivated} only, and the explicit pooling construction is what delivers the property.
Applied to the trained circuit, the same pooling reduces the equivariance error to order $10^{-8}$.
On the four trained PCAE2D hardware checkpoints, with an exact statevector pass over all eight transformed copies of each image, the mean error falls from $0.2777$ without pooling to $5.39\times10^{-8}$ with it.
This construction operates at the feature-grid level and serves here as a numerical diagnostic, rather than as a component of the trainable reconstruction pipeline.
Integrating it end-to-end (Section~\ref{sec:limitations}) is concrete future work.
This route to symmetry differs from equivariant quantum neural networks, in which every circuit layer is built to commute with the group action \citep{Nguyen2024EQNN}.
A recent image classifier of that kind builds $C_4$ rotation symmetry into the circuit itself \citep{SanSebastianSein2025}.
In PCAE the symmetry comes from averaging the observable grid over the group after measurement.
The circuit itself is not equivariant, and the same average applies equally to a classical feature map.

\begin{table}[t]
  \centering
  \caption{$D_4$ equivariance error for unpooled and pooled observable-grid maps.
  Lower is better.
  The pooled map is group-averaged over all eight square symmetries.
  Values are computed on the closed-form surrogate map.}
  \label{tab:d4_equivariance}
  \begin{tabular}{@{}l r r@{}}
    \toprule
    Model                    & Mean error              & Std.\ dev. \\
    \midrule
    $D_4$-pooled PCAE2D       & $9.57\times10^{-17}$    & $1.26\times10^{-17}$    \\
    No positional encoding   & $7.40\times10^{-1}$     & $1.52\times10^{-1}$     \\
    PCAE2D positional         & $8.15\times10^{-1}$     & $1.29\times10^{-1}$     \\
    Local/raw                & $8.37\times10^{-1}$     & $1.31\times10^{-1}$     \\
    \bottomrule
  \end{tabular}
\end{table}

\subsection{Pair-observable necessity in the pairwise-product protocol}
\label{sec:entanglement_necessity}

The diagnostic tests whether a pair-observable channel is needed to fit a target that depends on pair correlations.
The target is a fixed synthetic function of \emph{simulated} input variables, six per sample, and its pair terms are products of variables joined by edges of the $2\times3$ qubit grid.
Every control receives the same raw inputs as PCAE2D.
The protocol is audited for leakage.
For each feature block, we check whether a linear or near-linear map reaches near-zero residual on the features used to build the target, since such a fit would copy the target from the input.
The diagnostic reported here passes this check.
All feature maps in this diagnostic are closed-form, and no circuit is evaluated.
The pair-observable channel is represented by the analytic pair products that a pair measurement would supply.
By construction, these products match the pair terms of the target.
The target also has a term that is not a pair product, so no feature block reproduces it exactly.
The experiment demonstrates representational capability under a deliberately aligned protocol, not a general performance advantage.

\begin{figure}[t]
  \centering
  \includegraphics[width=0.9\textwidth]{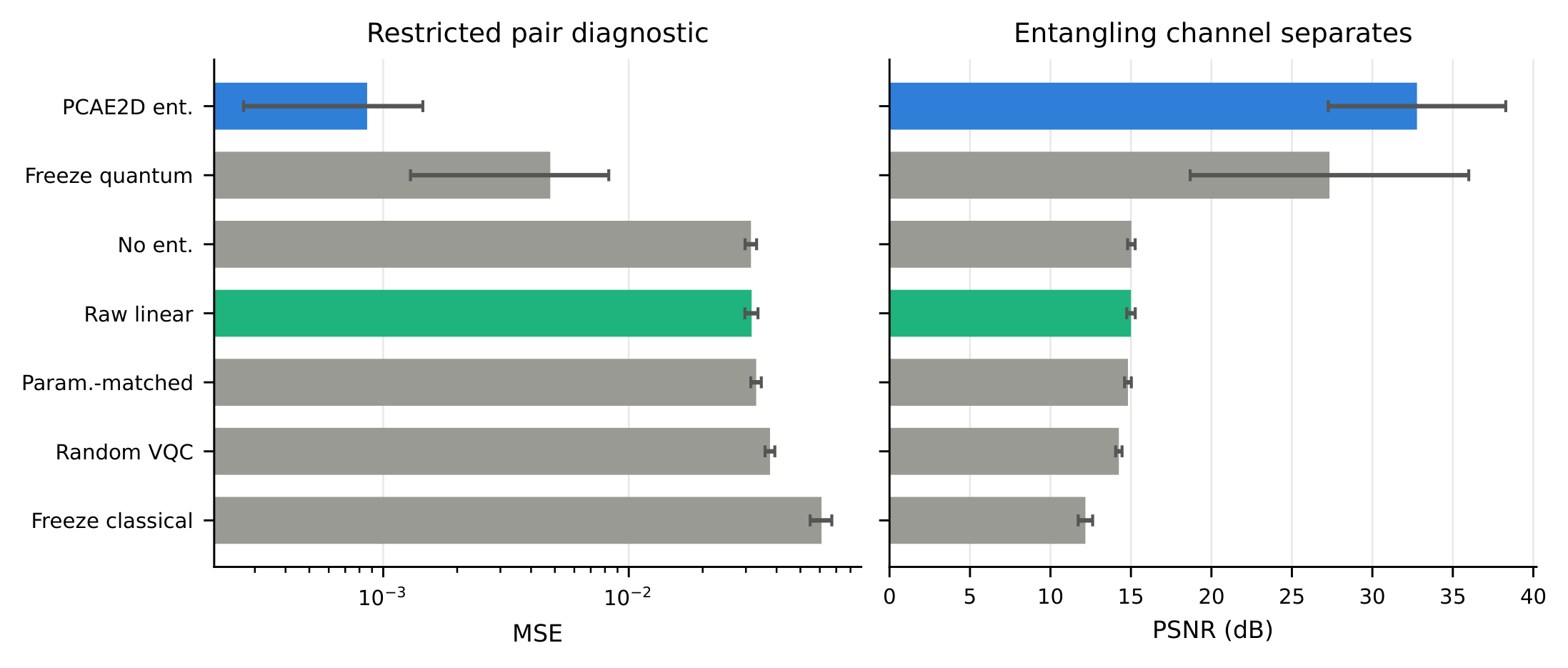}
  \caption[Restricted pair-correlation diagnostic summary]{Restricted pair-correlation diagnostic summary (mean $\pm$ std over five seeds).
  The pair-observable feature map dominates every control on this deliberately favourable task.}
  \label{fig:hvk_pair_diagnostic}
\end{figure}

Table~\ref{tab:hvk_pair_diagnostic} and Figure~\ref{fig:hvk_pair_diagnostic} report the result across five seeds.
PCAE2D's pair-observable channel reaches mean $R^2=0.9735$, or $32.77$~dB PSNR.
This is a $17.74$~dB improvement over a no-pair-observables control ($R^2=0.019$) and $18.52$~dB over the Random VQC control (Gaussian random latents, $R^2=-0.173$).
Randomly rescaling and masking the pair features (``freeze quantum''), with only the readout trained, still reaches $R^2=0.853$.
A raw-linear classical control ($R^2=0.0133$), a parameter-matched classical control ($R^2=-0.030$), and a frozen-classical control ($R^2=-0.903$) all fail on this deliberately pair-dependent target.
The no-pair-observables control retains only single-site nonlinear channels and removes every cross-site product feature.
Under the fixed linear readout, it therefore cannot construct the cross terms on which the target depends.

The mechanism is explicit under the fixed linear readout.
A linear map of single-qubit expectations $\{\langle O_i\rangle\}$ cannot form the cross-term $\langle O_i\rangle\langle O_j\rangle$ that the target depends on.
The pair measurement supplies the joint correlator $\langle O_iO_j\rangle$ directly as a feature.
Formally, let $z_i=\langle O_i\rangle$ and let every admissible predictor be affine, $f(z)=w^{\top}z+b$.
Then $\partial^2 f/\partial z_i\partial z_j=0$ for all $i,j$.
A target with a term $c\,z_iz_j$, $c\neq0$, has this mixed derivative equal to $c$ wherever $z_i$ and $z_j$ vary independently, so no affine $f$ can fit it.
Adding $z_iz_j$ to the feature vector makes the term linearly representable.
A pair-observable channel is therefore \emph{representationally necessary within the tested feature class and fixed linear-readout protocol} to fit this target.
A classical model that forms the same products, for example through quadratic features or a nonlinear readout, would also fit this target.
For this reason we scope the claim to the linear protocol rather than to classical models in general.

\begin{table}[t]
  \centering
  \caption[PCAE2D restricted pair-correlation diagnostic]{PCAE2D restricted pair-correlation diagnostic across five seeds.
  The task is deliberately favourable to pair-observable structure.
  The table gives evidence for pair-observable necessity under a matched linear readout.
  All features are closed-form.
  ``Random VQC'' is a block of Gaussian random features.}
  \label{tab:hvk_pair_diagnostic}
  \begin{tabular}{@{}l r r r@{}}
    \toprule
    Model                            & Mean MSE                  & Mean PSNR (dB) & Mean $R^2$ \\
    \midrule
    PCAE2D pair observables     & $8.5969\times10^{-4}$     & 32.77          & $ 0.9735$  \\
    Freeze quantum                   & $4.7904\times10^{-3}$     & 27.34          & $ 0.8533$  \\
    No pair observables & $3.1461\times10^{-2}$     & 15.03          & $ 0.0191$  \\
    Raw-linear classical             & $3.1654\times10^{-2}$     & 15.00          & $ 0.0133$  \\
    Parameter-matched classical      & $3.3033\times10^{-2}$     & 14.82          & $-0.0297$  \\
    Random VQC                       & $3.7616\times10^{-2}$     & 14.25          & $-0.1732$  \\
    Freeze classical                 & $6.0973\times10^{-2}$     & 12.17          & $-0.9025$  \\
    \bottomrule
  \end{tabular}
\end{table}

\subsection{A scoped training-dynamics diagnostic}
\label{sec:phase_transition}

Beyond the results above, we tracked two exploratory training-dynamics readouts.
These are a magnetisation-style statistic $M_z(t)$ and a signed energy-to-entanglement ratio $R_{ES}(t)$.
Here $M_z(t)$ is the mean of the local $\langle Z\rangle$ observables over all qubits and patches at optimisation step $t$.
$R_{ES}(t)$ is the learned Hamiltonian energy divided by the fixed sum of the MPS bond entropy features, so it tracks the energy up to a constant scale.
Both drift smoothly over training with run-to-run fluctuations, and neither shows a sharp transition.
Any apparent onset is a within-run median-plus-$2\sigma$ threshold crossing, rather than a calibrated statistical test.
The rule flags a run when the largest step-to-step change in $M_z$ exceeds that run's median change plus two standard deviations.

A negative control fixes the scope.
We applied the identical protocol to a classical-replacement variant, in which the Hamiltonian energy is identically zero throughout training.
That variant crosses the threshold in all $4/4$ tested cases, which is as often as the Hamiltonian-regularised model's $4/4$.
The threshold therefore fires whether the Hamiltonian term is present or absent, and it does not track that term.
We therefore report these traces as a descriptive readout on the learned energy and entanglement.

\section{Discussion}
\label{sec:discussion}

\subsection{Effect of the Hamiltonian energy term}
\label{sec:regularizer_inert}

In a single-seed comparison on \textit{Monalisa}, the reconstruction loss alone gives the higher PSNR.
Removing the signed energy term increases PSNR from $38.63$ to $44.56$~dB (Table~\ref{tab:hamiltonian_controls}).
The legacy objective adds a signed mean energy to the reconstruction loss.
The energy is linear in the learned couplings, which are not bounded, so it has no lower limit.
The optimiser can therefore lower the total loss by driving the energy down without improving reconstruction.
A contrastive variant, assigning lower energy to correct feature--position pairs than to shuffled pairs, recovers part of the gap ($42.92$~dB) but stays below its no-energy counterpart.
These single-seed values (\textit{Monalisa}, 200 steps, seed 42) are computed from the reconstruction term of the training loss at the final step, over the decoded patches and without seam blending.
The matched-budget ablation of Appendix~\ref{sec:core_multiseed_rerun} instead evaluates the stitched image after seam blending, which smooths a band around every patch boundary.
Blending lowers the PSNR of a well-fitted model, as for the hardware checkpoint, which gives $37.99$~dB without blending and $32.24$~dB with it (Appendix~\ref{sec:hardware_methodology}).
Under that evaluation, the no-energy-loss variant and the baseline agree within their seed-to-seed spread.
The two comparisons therefore use different evaluations and are not directly comparable.
The effect of the signed energy term is in general agreement with an independent study, which finds no benefit beyond simple classical regularisers from a VQE-inspired energy term added to the loss of a generative adversarial network \citep{Strnadel2025}.

The learned couplings do not enter the decoder directly, but the energy term contributes gradients during training and therefore changes the learned circuit representation.
It is consequently an auxiliary training objective as well as a post-training diagnostic.
A physically motivated regulariser has room to help when it resolves an otherwise underdetermined objective.
Under this protocol, the reconstruction loss alone provides the stronger optimisation signal.

\begin{table}[t]
  \centering
  \caption{Hamiltonian-sector controls (single-seed, \textit{Monalisa}, 200 steps, seed 42).
  PSNR is computed over the decoded patches, without seam blending.}
  \label{tab:hamiltonian_controls}
  \begin{tabularx}{\textwidth}{@{}L{0.30\textwidth} c Y@{}}
    \toprule
    Experiment                      & PSNR (dB) & Interpretation \\
    \midrule
    Legacy baseline, signed energy  & 38.63     & Standard energy-regularised run \\
    No energy loss                  & 44.56     & Removing energy improves PSNR \\
    Contrastive Hamiltonian core    & 42.92     & Stronger objective helps modestly
                                                  vs.\ baseline, still below
                                                  no-energy-loss \\
    \bottomrule
  \end{tabularx}
\end{table}

The training-dynamics readouts of Section~\ref{sec:phase_transition} provide a descriptive window into how the learned energy and entanglement evolve.

\subsection{Task-dependent representational scope}
\label{sec:dequant}
\label{sec:representativeness}

Whether a quantum feature map offers any advantage over a classical model is a property of the specific data and target, not a fixed property of the model class.
Dequantisation results have repeatedly shown that apparent quantum advantages on structured, low-entanglement classical data can be matched once the right classical feature is identified \citep{Huang2021PowerData, Tang2019}.
Where provable separations do exist, they are established for carefully constructed learning problems rather than for generic structured data \citep{Gyurik2023}.
We separate three questions.
First, PCAE at six qubits can be simulated exactly on a classical computer, which says little about larger scales.
Second, whether its model class admits an efficient classical replacement at scale is open.
Tensor networks are one route to such a replacement, and they have been used to simulate quantum circuits \citep{Patra2024IBM, Patra2025PEPS} and variational quantum algorithms \citep{Aizpurua2025}.
Tensor-network analyses also give conditions for such dequantisation of variational models \citep{Shin2024Dequant}, and a classical tie at small scale does not settle it.
Random Fourier features can also replace some variational models efficiently, under stated conditions \citep{Sweke2025}.
Quantum convolutional networks on common image benchmarks can be simulated efficiently by a classical method based on Pauli shadows \citep{Bermejo2026}.
Third, on held-out CIFAR-10, a closed-form surrogate of the PCAE2D map stays within 1~dB of matched classical feature maps.
This matches large benchmarks in which simple classical models often match or beat small quantum models, and removing entanglement often costs little \citep{Bowles2024}.
In PCAE the quantum processor evaluates only the circuit and returns the vector of measured Pauli expectation values.
The MPS compression is quantum-inspired but runs on classical hardware, as do the Fourier positional encoding, the $D_4$ pooling, the decoder and the optimisation.
Any quantum contribution to PCAE must therefore enter through the measured observable vector.

The restricted diagnostic of Section~\ref{sec:entanglement_necessity} is constructed so that the relevant structure is a product of two input variables, which raw or local features cannot represent under a fixed linear readout.
That is the regime in which PCAE's pair-observable channel separates.
Natural image patches at the resolution and patch size used here are low-entanglement, locally structured data \citep{Martyn2020, Lu2025}, where a linear or shallow classical readout is already close to sufficient.
For the images of Table~\ref{tab:sameset_multi_dataset}, the bipartite von Neumann entropy of the compressed patch states is at most $0.58$~nats for CIFAR-10 and $0.24$~nats for MedMNIST, over all five cuts of every patch.
The largest value allowed at bond dimension four is $\ln 4$.
The distinction lies in the measured pair-observable channel rather than in the circuit.
The two entangling layers do generate an entangled state, but whether the measured channel carries task-relevant information is a property of the target.
On natural-image targets the pair observables add little beyond the local ones.
On the constructed pairwise-product target they are necessary.
Together with the held-out comparisons of Section~\ref{sec:ablation_relationship}, these results delineate where the tested entangling representation does and does not yield a measurable benefit.

Three of PCAE's ingredients recur individually in the hybrid quantum-vision literature, covering MPS compression of structured input \citep{Stoudenmire2016, Han2018, Cirac2021, Guala2023}, a Pauli-correlator latent space handed to a classical decoder \citep{Havlicek2019}, and a physically motivated auxiliary regulariser alongside a task loss \citep{Shi2022}.
Its specific instantiation, a six-qubit VQC with a $19$--$27$-D correlator vector and a graph-Hamiltonian term, sits within the parameter ranges typical of that literature.
The measured behaviour is therefore evidence about the design pattern itself.
This boundary is established for the datasets, patch sizes and qubit count tested here.

\subsection{Qubit topology as an inductive bias}
\label{sec:topology}

The choice of qubit interaction graph (PCAE2D vs.\ PCAE1D) is an architectural inductive bias, and we test it directly under matched held-out conditions.
The real-circuit comparison (3 seeds, 3 held-out CIFAR-10 images) gives a PCAE1D-minus-PCAE2D mean PSNR difference of $0.16$~dB.
After averaging images within seed, the bootstrap 95\% CI is $[-0.38,0.46]$~dB and the seed-level Wilcoxon test gives $p=0.50$.
Both topologies fall about 1~dB below a matched random-latent control in this test (Appendix~\ref{app:random_latent}), so the comparison does not show what the qubit topology contributes to the latent.
PCAE2D completes the matched 90-step runs in roughly half the wall-clock time.
At the tested scale, topology is therefore primarily a resource-cost choice rather than a reliable accuracy advantage.
These tests do not establish formal equivalence.

Table~\ref{tab:differentiation} contrasts PCAE's structural properties against reference frameworks, including the three nearest hybrid-vision architecture families.
Its distinguishing features are a Pauli-correlator latent space, a 2D Fourier positional bias with enforced $D_4$ pooling, and a Hamiltonian energy term, each documented and scoped in the preceding sections.

\begin{table}[t]
  \centering
  \caption{Architectural differentiation from representative baseline designs, including the three nearest hybrid-vision architecture families (register-based QAE, quantum-kernel vision, TN-circuit classifier).}
  \label{tab:differentiation}
  \scriptsize
  \setlength{\tabcolsep}{3pt}
  \begin{tabularx}{\textwidth}{@{}L{0.13\textwidth} Y Y Y Y Y Y@{}}
    \toprule
    Axis
      & Conv.\ AE baseline
      & GAN baseline
      & Register-based QAE
      & Quantum-kernel vision
      & TN-circuit classifier
      & PCAE (ours) \\
    \midrule
    Latent representation
      & Continuous $\mathbb{R}^d$
      & Continuous $\mathbb{R}^d$
      & Quantum register
      & Scalar kernel value
      & MPS\slash circuit state
      & Pauli-correlator vector \\
    \addlinespace
    Spatial bias
      & Convolutional
      & Convolutional
      & Encoding-dependent
      & Encoding-dependent
      & MPS chain ordering
      & 2D Fourier encoding \\
    \addlinespace
    Auxiliary objective
      & Bottleneck / reconstruction
      & Adversarial loss
      & Fidelity / reconstruction
      & None (kernel value only)
      & Classification loss
      & Hamiltonian energy term \\
    \addlinespace
    Enforced group symmetry
      & Not in baseline
      & Not in baseline
      & Not in reference design
      & Not in reference design
      & Not in reference design
      & $D_4$ observable pooling \\
    \addlinespace
    Evaluated role
      & Reconstruction
      & Reconstruction
      & Compression
      & Classification
      & Classification
      & Reconstruction + diagnostics \\
    \bottomrule
  \end{tabularx}
\end{table}

\subsection{Validated scope}
\label{sec:limitations}

Every circuit-based reconstruction result in this paper is per-image trained.
The held-out comparison of Section~\ref{sec:ablation_relationship} uses a closed-form surrogate of the PCAE2D map.
On held-out images the trained circuit falls below a matched random-latent control at the tested budget, so its held-out generalisation is not established.
The per-image results of Section~\ref{sec:sameset_results} are not compared with a matched-size classical per-image network, as is usual for implicit neural representations.
A single-seed position-only ablation of the same decoder has a higher mean PSNR than PCAE2D on three of the six datasets (Table~\ref{tab:sameset_multi_dataset}).

The $D_4$-pooled map (Section~\ref{sec:d4}) is a feature-grid-level construction and numerical diagnostic, measured on the closed-form surrogate in Table~\ref{tab:d4_equivariance}.
Integrating it into the trainable end-to-end pipeline is future work.
The pair-observable necessity result (Section~\ref{sec:entanglement_necessity}) is scoped to a synthetic task with closed-form features, constructed to require products of input pairs.
Ordinary natural-image reconstruction lies outside that scope (see Section~\ref{sec:dequant}).
The training-dynamics traces (Section~\ref{sec:phase_transition}) are a descriptive readout.
The threshold fires whether the Hamiltonian term is present or absent, at 4/4 in both cases.

The hardware pilot (Section~\ref{sec:hardware_pilot}) is a five-image, unmitigated pilot with one run per image on the main backend rather than a statistically resolved hardware benchmark.
The matched real-circuit topology comparison uses 3 seeds and 3 held-out images per topology.
These results support a resource-cost interpretation rather than formal equivalence of the topologies.
Every result here is at a six-qubit, two-layer scale.
Trainability of variational models is known to degrade with width and depth through barren-plateau effects \citep{McClean2018, Larocca2025}.
Further, models that provably avoid barren plateaus are often also efficiently simulable on a classical computer \citep{Cerezo2025}.
Thus, whether the tested boundary moves at larger scale is an open question.

\section{Conclusion}
\label{sec:conclusion}

We presented the Pauli-Correlator Autoencoder (PCAE), a hybrid quantum--classical architecture for image reconstruction that assembles matrix-product-state patch features, a measured Pauli-correlator latent space and a learnable Hamiltonian energy term into a single pipeline (Section~\ref{sec:architecture}).
Across six real image datasets, per-image-trained PCAE2D models reach $25.8$--$41.6$~dB PSNR.
On held-out CIFAR-10, a closed-form surrogate of the PCAE2D observable map is equivalent within 1~dB to five resource-matched feature maps (Section~\ref{sec:ablation_relationship}).
A hardware pilot replayed trained checkpoints on IBM Quantum devices without retraining, establishing feasibility at a five-image scope with one run per image on the main backend (Section~\ref{sec:hardware_pilot}).

Pair observables are necessary on a synthetic task requiring products of input pairs, under a fixed linear readout (Section~\ref{sec:entanglement_necessity}).
The Hamiltonian energy term acts as an auxiliary training objective, and its learned energy is recorded as a descriptive readout during training (Section~\ref{sec:regularizer_inert}).
The pooled observable map is also exactly $D_4$-equivariant by construction, which we report as a design-correctness property rather than a quantum capability, since the same Reynolds averaging applied to a classical feature map is equally exact.
At the six-qubit scale tested here, PCAE fits varied real images and runs on real quantum hardware.

Future work should integrate the $D_4$-pooled map into an end-to-end trainable pipeline (Section~\ref{sec:limitations}), scale the hardware pilot toward a statistically resolved benchmark with error mitigation and repeated trials, test held-out generalisation of the trained circuit against the same resource-matched controls, search for further real-world tasks with pairwise-product structure where PCAE's pair-observable separation translates into a reconstruction advantage, and extend the resource-matched, leakage-audited evaluation protocol to other hybrid quantum vision architectures.

\backmatter

\bmhead{Acknowledgements}

We thank the maintainers of the open-source scientific computing libraries used in this work.
We acknowledge support from ANRF, India (grant ANRF/ECRG/2025/003752).

\section*{Statements and Declarations}

\begin{description}\itemsep4pt
  \item[Competing interests.] The authors declare no competing interests.

  \item[Funding.] This work was supported by the Anusandhan National Research Foundation (ANRF), India, under grant ANRF/ECRG/2025/003752.

  \item[Data availability.] Dataset provenance and access routes are cited in the manuscript.
  All datasets used (CIFAR-10, MNIST, Fashion-MNIST, and the PathMNIST, BloodMNIST and PneumoniaMNIST subsets of MedMNIST~v2) are publicly available.

  \item[Code availability.] The code supporting this study is available from the corresponding author upon reasonable request.

  \item[Author contributions.]
  S.C. and S.P. conceived the project.
  S.C. wrote the code and carried out the experiments and analysis.
  S.P. supervised the work.
  Both authors wrote the manuscript.
\end{description}

\begin{appendices}
\renewcommand{\theHtable}{app.\thetable}%
\renewcommand{\theHfigure}{app.\thefigure}%
\renewcommand{\theHequation}{app.\theequation}%

\section{Training settings and evaluation protocol}
\label{sec:training_settings}

All trained models use Adam with full-batch updates over the patches of their training images.
The loss is the reconstruction mean squared error plus $\lambda=0.01$ times the mean Hamiltonian energy, except where an ablation removes or replaces the energy term.
The held-out comparison fits no network.
Its linear readout is solved in closed form as a ridge regression with regularisation $10^{-6}$.
Table~\ref{tab:training_settings} lists the settings of each experiment.

\begin{table}[!htbp]
\centering
\caption{Training settings.
Learning rates are for Adam.
Patches are non-overlapping unless a stride is given.}
\label{tab:training_settings}
\footnotesize
\setlength{\tabcolsep}{4pt}
\begin{tabular}{@{}l l l r l l@{}}
\toprule
Experiment & Model & Patches & Steps & Learning rate & Seeds \\
\midrule
\textit{Monalisa} hardware checkpoint$^{a}$ & PCAE1D & 16, $64\times64$ & 200 & 0.003 & 42 \\
240-step ablation & PCAE1D & 16, $64\times64$ & 240 & 0.003 & 0, 1, 2$^{b}$ \\
Six-dataset fits & PCAE2D & 16, $8\times8$ & 100 & 0.004 & not recorded \\
CIFAR-10 hardware fits & PCAE2D & 16, $8\times8$ & 200 & 0.004 & not recorded \\
Topology comparison & 1D / 2D & 49, $8\times8$, stride 4 & 90 & 0.003 / 0.004 & 0, 1, 2 \\
Held-out comparison & surrogate & 16, $8\times8$ & n/a & n/a (ridge) & 0--4$^{c}$ \\
\bottomrule
\end{tabular}

\smallskip
\raggedright $^{a}$\,The hardware pilot uses the legacy \texttt{shared-baseline-seed-42} checkpoint. The Hamiltonian-control values in Table~\ref{tab:hamiltonian_controls} are independent fresh reruns made with the same nominal 100-step, seed-42 settings; they are not the hardware checkpoint.
\\$^{b}$\,The reported baseline mean also includes seed 3.
\\$^{c}$\,Split seeds; each split fits the ridge readout on six CIFAR-10 images and holds out four disjoint images.
\end{table}

Where Table~\ref{tab:training_settings} lists seeds, the experiment drivers set the Python, NumPy and PyTorch random seeds explicitly.
The drivers of the six-dataset fits and the CIFAR-10 hardware fits did not set or record a seed, so these are reported as not recorded.
``Mean $\pm$ std'' denotes the mean and the population standard deviation over the unit named in each caption.
Bootstrap intervals are percentile 95\% intervals, and Wilcoxon values are two-sided signed-rank tests.
The runs used PyTorch 2.6.0 with CUDA 12.4, PennyLane 0.42.3 and quimb 1.14.0 on an NVIDIA GeForce GTX 1050.

\section{Matched-budget component ablation}
\label{sec:core_multiseed_rerun}

All nine variants were trained on \textit{Monalisa} with a matched 240-step budget and an identical learning rate and schedule, for three seeds each (seeds 0--2), with a fourth seed (3) for the baseline (Table~\ref{tab:core_multiseed_240}).
The first freeze-quantum attempts at seeds 0--2 failed with undefined (NaN) metrics because of an implementation error, and only the three successful reruns enter its mean.
Freeze-classical trains only the circuit and the Hamiltonian couplings, with the decoder and projection layers fixed, and freeze-quantum does the reverse.
No-entanglement removes the CNOT rings, and no-energy-loss removes the energy term.
No-MPS replaces the MPS feature extractor with a 46-dimensional vector of direct patch statistics (mean, standard deviation, minimum, maximum, seven percentiles and a normalised intensity histogram), and keeps the projection, circuit, positional encoding and decoder.
Random-VQC replaces the circuit observables at every forward pass with an independently resampled standard-Gaussian vector of the same dimension, sets the energy to zero, and trains the decoder normally.
Classical-replacement replaces the circuit with a tanh map, and classical-matched limits that map to the circuit's parameter count.
The classical-replacement variant exceeds the quantum baseline by $0.081$~dB, against a pooled standard deviation of $0.019$~dB.
The classical-matched variant falls $0.197$~dB below the baseline.
These differences are measured on the stitched image after seam blending, which compresses the differences between well-fitted variants (Section~\ref{sec:regularizer_inert}).

\begin{table}[!htbp]
\centering
\caption{Matched-budget (240-step) component ablation on \textit{Monalisa}, three seeds per variant (four for the baseline), mean $\pm$ std.
PSNR and SSIM are computed on the stitched image after seam blending.}
\label{tab:core_multiseed_240}
\footnotesize
\setlength{\tabcolsep}{4pt}
\begin{tabular}{@{}lcc@{\hspace{14pt}}lcc@{}}
\toprule
Variant & PSNR (dB) & SSIM & Variant & PSNR (dB) & SSIM \\
\midrule
Baseline (quantum) & $33.42\pm0.02$ & $0.99385\pm0.00003$ & Classical-replacement & $33.50\pm0.01$ & $0.99395\pm0.00002$ \\
Freeze-quantum & $33.39\pm0.06$ & $0.99381\pm0.00006$ & Classical-matched & $33.22\pm0.19$ & $0.99356\pm0.00027$ \\
No-entanglement & $33.48\pm0.02$ & $0.99392\pm0.00003$ & Random-VQC & $28.09\pm0.23$ & $0.97907\pm0.00135$ \\
No-MPS & $33.34\pm0.05$ & $0.99374\pm0.00008$ & Freeze-classical & $10.94\pm0.00$ & $0.0219\pm0.0001$ \\
No-energy-loss & $33.44\pm0.01$ & $0.99387\pm0.00001$ & & & \\
\bottomrule
\end{tabular}
\end{table}

\section{Held-out reconstruction}
\label{app:multidataset}

The held-out protocol of Section~\ref{sec:ablation_relationship} was repeated on cached subsets of six datasets, with three stratified split seeds (Table~\ref{tab:multi_dataset_reconstruction}).
From the first 400 images of each dataset, each split draws 20 training and 10 held-out images per class (one and one for CIFAR-10), and skips any class with fewer than 30 images among them.
This gives 10/10, 200/100, 200/100, 160/80, 120/60 and 40/20 training/held-out images per split for CIFAR-10, MNIST, Fashion-MNIST, PathMNIST, BloodMNIST and PneumoniaMNIST, so PathMNIST uses 8 of its 9 classes and BloodMNIST 6 of its 8.
These results predate the correction of the positional channel in the CIFAR-10 comparison of Section~\ref{sec:ablation_relationship}.
The earlier width selection retained only two of the eight positional features for the PCAE2D surrogate and none for the quadratic control.
The no-entanglement and raw/local controls retained all eight.
The feature allocation here therefore differs from that of the corrected comparison.
The local/raw control has equal or higher mean PSNR on every dataset.
After averaging within each split seed, all six seed-bootstrap intervals exclude zero, and each two-sided signed-rank test gives $p=0.25$, the smallest value attainable with three seeds.

\begin{table}[!htbp]
\centering
\caption{Held-out PSNR (dB) on cached subsets, mean $\pm$ std over held-out images, for the best local/raw control and the PCAE2D surrogate.}
\label{tab:multi_dataset_reconstruction}
\footnotesize
\begin{tabular}{lccc}
\toprule
Dataset & Best local/raw & PCAE2D surrogate & Higher mean \\
\midrule
CIFAR-10 & $21.17{\pm}1.98$ & $19.99{\pm}2.02$ & Local/raw \\
MNIST & $16.55{\pm}1.27$ & $16.12{\pm}1.34$ & No-entanglement/local \\
Fashion-MNIST & $17.67{\pm}1.71$ & $17.33{\pm}1.80$ & Local/raw \\
PathMNIST & $27.52{\pm}6.02$ & $27.01{\pm}5.81$ & Local/raw \\
BloodMNIST & $24.79{\pm}1.78$ & $24.24{\pm}1.66$ & Local/raw \\
PneumoniaMNIST & $26.49{\pm}2.48$ & $25.73{\pm}2.29$ & Local/raw \\
\bottomrule
\end{tabular}
\end{table}

\label{app:random_latent}%
In the held-out topology test of Section~\ref{sec:topology}, PCAE1D reaches $11.73\pm1.99$~dB and PCAE2D reaches $11.57\pm1.53$~dB.
A random-latent control under the identical protocol replaces the observable vector with resampled noise and trains the decoder normally.
It reaches $12.74\pm2.21$~dB for PCAE1D and $12.64\pm2.12$~dB for PCAE2D.
The trained circuits are about 1~dB below this floor, and the paired difference is negative in all six seed-level comparisons (two-sided Wilcoxon $p=0.031$).
At 90 steps, only the PCAE1D-minus-PCAE2D difference in this test is informative.

In a separate end-to-end test, one PCAE2D model is trained with full-batch Adam for 20 epochs on 150 CIFAR-10 images and evaluated, without further training, on 50 held-out images.
Each of three seeds reshuffles the split.
The model reaches $14.39\pm0.26$~dB, against $15.61\pm0.29$~dB for a parameter-matched classical linear control trained the same way.
The control is higher at each of the three seeds.

\section{Hardware pilot methodology}
\label{sec:hardware_methodology}

The \textit{Monalisa} circuit is the legacy PCAE1D checkpoint \texttt{shared-baseline-seed-42}, trained for 200 steps with seed 42 and learning rate $0.003$.
It is distinct from the fresh ``Legacy baseline, signed energy'' rerun of Table~\ref{tab:hamiltonian_controls}.
Its noiseless reference of $37.99$~dB (Table~\ref{tab:hardware_pilot_summary}) is an exact statevector replay of the saved checkpoint through the hardware-reconstruction pipeline.
Both the hardware and the simulator PSNR are computed over the decoded patches, without stitching them or blending the seams.
The training-run evaluation of the same checkpoint blends the seams of the stitched image and records $32.24$~dB.
The four CIFAR-10 circuits are PCAE2D checkpoints, each trained on its own image with 16 non-overlapping $8\times8$ patches (Table~\ref{tab:training_settings}).
PCAE1D measures $Z$, $X$, $ZZ$, $XX$ and $YY$ (27 observables).
Its pair correlators are joint parities of the same all-$Z$, all-$X$ and all-$Y$ basis measurements that give the single-qubit values, so three measurement circuits per patch suffice, or 48 circuits for \textit{Monalisa}.
PCAE2D measures $Z$, $X$ and $ZZ$ (19 observables) and needs two circuits per patch, or 128 circuits for the four CIFAR-10 images.

Before any hardware submission, every circuit was checked against the cached training-time observables by exact statevector simulation.
The largest absolute error was $7.89\times10^{-4}$ for \textit{Monalisa} and at most $9.42\times10^{-6}$ for the CIFAR-10 checkpoints.
On hardware, the mean absolute deviation of the observables from the statevector values was $0.096$--$0.125$, so gate translation is not the main source of the hardware--simulator PSNR gap.
Hardware counts were converted to expectation values with the standard parity estimator and passed, with the training-time positional encoding, through the unmodified trained decoder.

\end{appendices}

\bibliography{sn-bibliography}

\end{document}